\documentclass[conference]{IEEEtran}
\usepackage{array}

\usepackage{multirow}
\usepackage{tabulary}

\ifCLASSINFOpdf
  \usepackage[pdftex]{graphicx}
  \graphicspath{{figs/}}
  \DeclareGraphicsExtensions{.pdf,.jpeg,.png}
\else
  \usepackage[dvips]{graphicx}
  \graphicspath{{figs/}}
  \DeclareGraphicsExtensions{.eps}
\fi
\usepackage[cmex10]{amsmath}
\usepackage{amssymb}
\usepackage{amsbsy}
\usepackage{caption}
\usepackage[caption=false,font=footnotesize]{subfig}
\usepackage{fixltx2e}
\usepackage{url}

\newcommand{\wildcard}{{\boldsymbol ?}}
\newcommand{\gap}{{\boldsymbol \star}}
\newlength{\aligncharw}
\newcommand{\agap}{\makebox[\aligncharw][s]{\scriptsize{$\gap$}}}
\newcommand{\eg}{e.g.\,}
\newcommand{\ie}{i.e.\,}
\newcommand{\cf}{cf.\,}

\begin{document}
%
\title{From Network Traces to System Responses: Opaquely Emulating Software Services}



%
\author{\IEEEauthorblockN{Miao Du\IEEEauthorrefmark{1},
Steve Versteeg\IEEEauthorrefmark{2},
Jean-Guy Schneider\IEEEauthorrefmark{1},
John Grundy\IEEEauthorrefmark{1},
Jun Han\IEEEauthorrefmark{1}}
\IEEEauthorblockA{\IEEEauthorrefmark{1}School of Software and Electrical Engineering\\
Swinburne University of Technology,
Hawthorn, Victoria 3122, Australia\\
Email: \{miaodu,jschneider,jgrundy,jhan\}@swin.edu.au}
\IEEEauthorblockA{\IEEEauthorrefmark{2}CA Technologies, Melbourne, Victoria 3004, Australia\\
Email: steve.versteeg@ca.com}}


\maketitle

\begin{abstract}
\boldmath
Enterprise software systems make complex interactions with other services in their environment. Developing and testing for production-like conditions is therefore a challenging task. Prior approaches include emulations of the dependency services using either explicit modelling or record-and-replay approaches. Models require deep knowledge of the target services while record-and-replay is limited in accuracy. We present a new technique that improves the accuracy of record-and-replay approaches, without requiring prior knowledge of the services. The approach uses multiple sequence alignment to derive message prototypes from recorded system interactions and a scheme to match incoming request messages against message prototypes to generate response messages. We introduce a modified Needleman-Wunsch algorithm for distance calculation during message matching, wildcards in message prototypes for high variability sections, and entropy-based weightings in distance calculations for increased accuracy. Combined, our new approach has shown greater than 99\% accuracy for four evaluated enterprise system messaging protocols.
\end{abstract}


%
\IEEEpeerreviewmaketitle

\section{Introduction}

Software systems are becoming ever increasing inter-connected. Service emulation supports developing and testing a software system, independent of the other systems on which it depends \cite{distributedsystemeval}. In a typical deployment scenario, an enterprise system might interact with many other systems, such as a mainframe, directory servers, databases and other types of software services. In order to check whether the enterprise system will function correctly in terms of its interactions with these services, it is necessary to test it in as realistic an environment as possible, prior to the actual deployment.  Getting access to the actual production environment for testing is not possible due to the risk of disruption.  Large organisations often have a test environment, which is a close replication of their production environment, but this is very expensive.  Furthermore the test environment is in high demand, so software developers will have only limited access to it.  Enabling developers to have continuous access to production-like conditions to test their application is an important part of Development-Operations (known as `DevOps')~\cite{erich2014report}\cite{bass:2015}.

One popular approach that developers use to test their application's dependence on other systems is to install the other systems on virtual machines (such as VMware) \cite{Sugerman:01}.  However virtual machines are time consuming to configure and maintain.  Furthermore the configuration of the systems running on the virtual machine is likely to be different to the production environment. An alternative that is gaining traction is service emulation, where models of services are emulated - sometimes into the many thousands of service instances - to provide more realistic scale and less complicated configuration \cite{hine:09a}. However, existing approaches to service emulation rely on system experts explicitly modelling the target services and this requires very detailed knowledge of message protocols and structure.  This is often infeasible if the required knowledge is unavailable for the wide range of services, especially legacy services, in a real deployment environment \cite{Ghosh99issuesin}, and is besides very time consuming and error-prone.

Our aim is to develop an automated approach to service emulation which uses no explicit knowledge of the services, their message protocols and structures and yet can simulate -- to a high degree of accuracy and scale -- a realistic enterprise system deployment environment.  To achieve this we need an automated, accurate, efficient and robust method for service emulation that derives service responses from collected actual network traces between an enterprise system and a service -- the focus of this paper.

Our previous work \cite{Du:2013} relied on searching the collected message traces to automatically generate service responses.  Unfortunately this was limited to text based protocols and became inefficient for large message traces.  The method we employed was sensitive to mismatching by payload versus operation type (\ie was not robust.)  Our subsequent work \cite{Du:2013SoftMine} used clustering to improve efficiency but decreased accuracy.  The aim of the research described in this paper is to achieve both higher accuracy and efficiency through the use of descriptive and robust message prototypes which can be applied to both binary and text protocols. Our approach uses:

\begin{itemize}
    \item a multiple sequence alignment algorithm to derive message prototypes,
    \item wildcards in message prototypes for the sections with high variability,
    \item a modified Needleman-Wunsch algorithm for message distance calculations, and
    \item entropy-based weightings in distance calculations for increased accuracy.
\end{itemize}

We present a set of experiments with four enterprise system messaging protocols, including binary and textual LDAP, a binary Mainframe protocol, a textual JSON (Twitter) and SOAP services. These experiments show overall a greater than $99\%$ accuracy in the generated response messages for the four protocols tested. Additionally they show efficient performance in generating the emulated service responses, enabling scaling within an emulated deployment environment.


\section{Motivation and Related Work}
\label{sec:relatedwork}

Consider an example enterprise software system which is to be integrated with other systems in its production environment as part of its operation.  The other systems (called services) include a legacy mainframe program, a directory server and a web service.  The behaviour of the example enterprise software system depends on the responses it receives from these dependency services.  The directory server uses a proprietary protocol which is poorly documented.
It has operation types add, search, modify and delete, each with various payloads and response messages. The operation type is encoded in a single character. A small sample set of request and response messages (a \emph{transaction library}) for the directory service is shown in Table~\ref{tab:tl}.  Such message traces are captured by either network monitoring tools, such as Wireshark \cite{wireshark}, or proxies. In the following sections we use this sample transaction library to illustrate how our method works. This transaction library is from a fictional protocol that has some similarities to the widely used LDAP protocol~\cite{ldap}, but is simplified to make our running example easier to follow.

The enterprise system under test needs to interact with this service, to search for user identities, register new identities, de-register identities and so on. Other systems in the environment share the same service. For complex deployment environments we want to extensively test the enterprise system as to its protocol conformance, robustness and scaling.

\begin{table*}[t]
\begin{center}
\begin{tabular}{|c||l|l|}
\hline
Index & Request & Response \\ \hline\hline
1 & \{id:1,op:S,sn:Du\} & \{id:1,op:SearchRsp,result:Ok,gn:Miao,sn:Du,mobile:5362634\} \\ \hline
13 & \{id:13,op:S,sn:Versteeg\} & \{id:13,op:SearchRsp,result:Ok,gn:Steve,sn:Versteeg,mobile:9374723\} \\ \hline
24 & \{id:24,op:A,sn:Schneider,mobile:123456\} & \{id:24,op:AddRsp,result:Ok\} \\ \hline
275& \{id:275,op:S,sn:Han\} & \{id:275,op:SearchRsp,result:Ok,gn:Jun,sn:Han,mobile:33333333\} \\ \hline
490 & \{id:490,op:S,sn:Grundy\} &
\{id:490,op:SearchRsp,result:Ok,gn:John,sn:Grundy,mobile:44444444\} \\ \hline
2273 & \{id:2273,op:S,sn:Schneider\} & \{id:2273,op:SearchRsp,result:Ok,sn:Schneider,mobile:123456\} \\ \hline
2487 & \{id:2487,op:A,sn:Will\} & \{id:2487,op:AddRsp,result:Ok\} \\ \hline
3106 & \{id:3106,op:A,sn:Hine,gn:Cam,Postcode:33589\} & \{id:3106,op:AddRsp,result:Ok\} \\
\hline
\end{tabular}
\end{center}
\caption{Directory Service Interaction Library Example}
\label{tab:tl}
\end{table*}

The most common approach to providing a testing environment for enterprise systems using such a messaging protocol is to use virtual machines \cite{li:10}. Implementations of the services are deployed on virtual machines and communicated with by the system under test. Major challenges with this approach include configuration complexity \cite{vm4testing} 
 and the need to maintain instances of each and every service type in multiple configurations \cite{grundy:05}. Recently, cloud-based testing environments have emerged to mitigate some of these issues \cite{testenvcloud}.

Emulated testing environments for enterprise systems, relying on service models, are another approach. When sent messages by the enterprise system under test, these respond with approximations of ``real'' service response messages \cite{servicemodel}. Kaluta \cite{hine:thesis} is proposed to provision emulated testing environments. Challenges with these approaches include developing the models, lack of precision in the models, especially for complex protocols, and ensuring robustness of the models under diverse loading conditions \cite{sun2012usefulness}.  To assist developing reusable service models, approaches either reverse engineer message structures \cite{messageformatinference}\cite{cui:07a}\cite{wang2012semantics}, or discover \cite{processmining}\cite{processdiagnostics} processes for building behavioral models \cite{behavioralmodel}. While these allow engineers to develop more precise models, none of them can automate interaction between enterprise system under test and the emulated dependency services.

Recording and replaying message traces is an alternative approach. This involves recording request messages sent by the enterprise system under test to real services and response messages from these services, and then using these message traces to `mimic' the real service response messages in the emulation environment \cite{cui:06}. Some approaches combine record-and-replay with reverse-engineered service models. CA Service Virtualization~\cite{Michelsen:11} is a commercial software tool, which can emulate the behaviour of services. The tool assumes knowledge of protocol message structures to model services and mimic interactions automatically.

%
%
%
%

\section{Approach}
\label{sec:approach}

\begin{figure}[h]
     \centering
     \includegraphics[width=8.6cm]{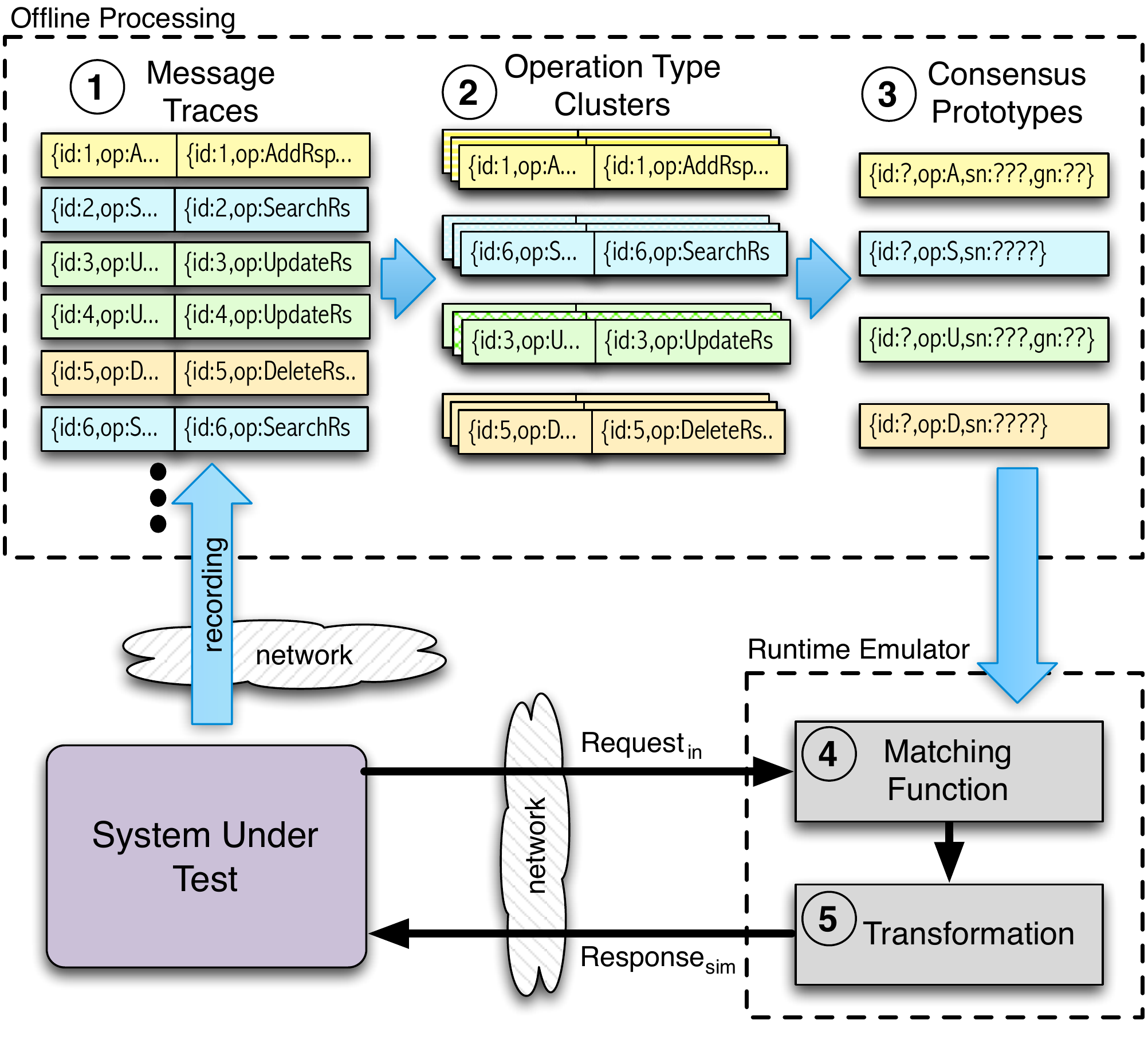}
     \caption{System overview}
     \label{fig:system_overview}
\end{figure}

Our goal is to produce an emulation environment for enterprise system testing that uses message trace recordings collected a priori to produce a response on behalf of a service when 
 invoked by a system-under-test at runtime.

The general approach is to cluster the trace recordings
into groups of similar messages and then
formulate a single representation for the
request messages of each cluster.
This accelerates runtime performance by enabling incoming requests from
the system under test to be compared only to the cluster
representations, rather than the entire transaction library.  
Previous work \cite{Du:2013SoftMine} selected the cluster \emph{centroid}\footnote{
The centroid is the transaction with the minimised total distance
from the other transactions in the cluster.}
request as the representative.
However accuracy decreased as the information from the
other requests in the cluster was discarded. Our current aim is to improve accuracy,
but preserve efficiency, by generating 
\emph{prototypes} which capture the common features of the range of
requests in each cluster.

The approach for generating and using our prototypes (depicted in figure~\ref{fig:system_overview}) has the following steps:


\begin{enumerate}
\item We collect message traces of communications between a client and the real target service and store them in \emph{a transaction library} with a transaction being a request-response pair.
\item We then cluster the transaction library, with the goal of grouping transactions by operation type.  We do not consider the state of service under which the requests are issued. (This may give lower accuracy but is still useful in many cases as discussed in Section~\ref{sec:discussion}.)
\item We then derive a \emph{request consensus prototype} (a message pattern) for each operation type cluster by:
\begin{enumerate}
    \item Aligning all request messages of each cluster to reveal their common features for each operation type by inserting gap characters;
	\item Extracting a consensus character sequence from the aligned request messages (i.e., the request message prototype) by selecting or deleting a character at each character position;
    \item Calculating positional weightings to prioritise different sections of each request consensus prototype. 
\end{enumerate}
\item At runtime, for an incoming request message, we use \emph{a matching distance calculation} technique to select the nearest matching request consensus prototype.
\item We then perform dynamic substitutions on a specifically chosen response message from the identified operation type cluster to generate a final response message to be sent back to the enterprise system under test.
\end{enumerate}

Steps 1 to 3 are performed offline to prepare request consensus prototypes.
Steps 4 and 5 are performed at runtime to generate live responses to the system under test messages.


\subsection{Needleman-Wunsch}
\label{ss:nw}
The Needleman-Wunsch algorithm \cite{needleman:1970} is used at different
steps during the offline and runtime processing.
It is a dynamic programming algorithm which finds the
globally optimal alignment for two sequences of symbols in $O(MN)$ time, where
$M$ and $N$ are the lengths of the sequences.  Needleman-Wunsch
uses a progressive scoring function $S$, which gives an incremental score for each pair
of symbols in the alignment.  A different score will be given depending
on whether the symbols are identical, different or a gap has been inserted.

\subsection{Collecting and Clustering Interaction Messages}
\label{sec:clustering}

The first step is to record real message exchanges between a client
(such as a previous version of the system under test) and the service we aim to emulate.  A transaction is a
request-response pair of the communications, where the service may
respond with zero or more response messages to a
request message from the client.  We record requests and responses at the network level
(using a tool such as Wireshark), recording the bytes in TCP packet
payloads. This makes no assumptions
about the message format of the service application.

Having recorded our transaction library, the next step is to group the
transactions by operation type, but again without assuming any
knowledge of the message format.  To achieve this we use a distance function-based clustering technique.
In our previous work \cite{Du:2013SoftMine} we considered multiple cluster distance functions and found that the response similarity (as measured by edit distance, calculated using the Needleman-Wunsch algorithm \cite{needleman:1970}) was the most effective method for grouping transactions
of the same operation type.  We therefore use the same technique in this work, grouping
transactions by the similarity of their response messages.  The clustering algorithm used was VAT
\cite{VAT}, one among many alternatives, as we found it to be effective in our previous work \cite{Du:2013SoftMine}.


Applying our clustering technique to the example transaction library
yields two clusters, as shown in tables \ref{tab:searchcluster} and
\ref{tab:addcluster}, corresponding to search operations and add
operations, respectively.

%

\begin{table*}[h]
\begin{center}
\begin{tabular}{|c||l|l|}
\hline
Index & Request & Response \\ \hline\hline
1 & \{id:1,op:S,sn:Du\} & \{id:1,op:SearchRsp,result:Ok,gn:Miao,sn:Du,mobile:5362634\} \\ \hline
13 & \{id:13,op:S,sn:Versteeg\} & \{id:13,op:SearchRsp,result:Ok,gn:Steve,sn:Versteeg,mobile:9374723\} \\ \hline
275& \{id:275,op:S,sn:Han\} & \{id:275,op:SearchRsp,result:Ok,gn:Jun,sn:Han,mobile:33333333\} \\ \hline
490 & \{id:490,op:S,sn:Grundy\} &
\{id:490,op:SearchRsp,result:Ok,gn:John,sn:Grundy,mobile:44444444\} \\ \hline
2273 & \{id:2273,op:S,sn:Schneider\} & \{id:2273,op:SearchRsp,result:Ok,sn:Schneider,mobile:123456\} \\ \hline
\end{tabular}
\end{center}
\caption{Cluster 1 (search operations)}
\label{tab:searchcluster}
\end{table*}

\begin{table*}[!h]
\begin{center}
\begin{tabular}{|c||l|l|}
\hline
Index & Request & Response \\ \hline\hline
24 & \{id:24,op:A,sn:Schneider,mobile:123456\} & \{id:24,op:AddRsp,result:Ok\} \\ \hline
2487 & \{id:2487,op:A,sn:Will\} & \{id:2487,op:AddRsp,result:Ok\} \\ \hline
3106 & \{id:1106,op:A,sn:Hine,gn:Cam,postalCode:33589\} & \{Id:1106,Msg:AddRsp,result:Ok\} \\
\hline
\end{tabular}
\end{center}
\caption{Cluster 2 (add operations)}
\label{tab:addcluster}
\end{table*}

\subsection{Aligning Request Messages}


To formulate a representative cluster prototype, we first align the request messages in a cluster to determine the common features of these request messages. Then we extract the common features while accommodating variations. In aligning the request messages of a cluster, we adopt the multiple sequence alignment (MSA) technique originated from Biology. MSA was first used to align three or more biological sequences to reveal their structural commonalities~\cite{book:biologicalsequenceanalysis}\footnote{For this reason this technique has
also been widely used to reverse-engineer protocol message
structures~\cite{Comparetti:2009,beddoe:2004}.}.
Using MSA for
revealing commonalities of interaction messages offers a number of advantages over other techniques, such as $n$-grams~\cite{ngram:wang2012semantics}. Firstly, it is able to \emph{effectively} handle messaging protocols with single-byte operation fields (\eg in the binary LDAP protocol) as well as multi-byte operation fields. Secondly, it does not require a predetermined matching sequence width (\eg, $n$ in $n$-gram).  

\label{ss:msa}

In particular, we adopt \emph{ClustalW}\cite{clustalw}, a widely used heuristic technique
for MSA.  It is memory efficient and is shown to produce high accuracy
alignments in polynomial computation time for empirical datasets (in contrast to the original NP-complete MSA technique~\cite{wang1994complexity}).  
ClustalW is a progressive MSA algorithm, where pairwise sequence
alignment results are iteratively integrated into the multiple
sequence alignment result.  An overview of the ClustalW algorithm is
as follows:
\begin{enumerate}
   \item All $N(N-1)/2$ pairs of sequences are aligned to calculate their similarity ratio by using the Needleman-Wunsch algorithm, where $N$ is the number of sequences.
   \item A $N \times N$ distance matrix is built for capturing distance calculation.
   \item A \emph{guide tree} is constructed from the distance matrix by applying a neighbour-joining clustering algorithm.  (A guide tree is a tree data structure which organises the similarities between sequences.)
   \item The guide tree is used to guide a progressive alignment of sequences from the leaves to the root of the tree.
 \end{enumerate}

Figure~\ref{chap7fig:alignment} shows the multiple sequence alignment
results of applying the $ClustalW$ algorithm to the example clusters from tables
\ref{tab:searchcluster} and \ref{tab:addcluster}.  The MSA results are
known as \emph{profiles}.
Gaps which were inserted during the alignment
process are denoted with the `$\gap$' symbol.
Note that the common sequences for the requests in each cluster have now
been aligned.

\newcommand{\alignedSrqs}{
	\begin{minipage}[b]{.45\linewidth}
	\texttt{\footnotesize{
	\{id:\agap{\agap}1{\agap},op:S,sn:{\agap}{\agap}{\agap}{\agap}{\agap}{\agap}{\agap}Du\} \linebreak
	\{id:{\agap}{\agap}13,op:S,sn:{\agap}Versteeg\} \newline
	\{id:2273,op:S,sn:Schneider\} \newline
	\{id:275{\agap},op:S,sn:{\agap}Han{\agap}{\agap}{\agap}{\agap}{\agap}\} \newline
	\{id:490{\agap},op:S,sn:Grundy{\agap}{\agap}{\agap}\} \newline
	}}
	\end{minipage}
}

\newcommand{\alignedArqs}{
	\begin{minipage}[b]{.45\linewidth}
	\texttt{\footnotesize{
	\{id:24{\agap}{\agap},op:A,sn:Schne{\agap}{\agap}{\agap}{\agap}{\agap}ider{\agap}{\agap},mo{\agap}bil{\agap}{\agap}{\agap}e:123456\}
	\{id:2487,op:A,sn:W{\agap}{\agap}{\agap}{\agap}{\agap}{\agap}{\agap}{\agap}{\agap}{\agap}{\agap}{\agap}{\agap}{\agap}{\agap}{\agap}{\agap}{\agap}{\agap}{\agap}il{\agap}{\agap}{\agap}l{\agap}{\agap}{\agap}{\agap}{\agap}{\agap}{\agap}\}
	\{id:3106,op:A,sn:Hi{\agap}ne,gn:Cameron,postalCode:3{\agap}3589\}
	}}
	\end{minipage}
}

\begin{figure}[ht]%
	\subfloat[Cluster 1 alignment \label{chap7fig:reqalignment}]{\alignedSrqs}%
     \qquad \newline
	\subfloat[Cluster 2 alignment \label{chap7fig:reqalignment}]{\alignedArqs}
\caption{MSA results of the requests in tables \ref{tab:searchcluster} and \ref{tab:addcluster}}%
\label{chap7fig:alignment}%
\end{figure}

\subsection{Formulating the Request Consensus Prototype}
\label{subsec:consensusseqcal}

Having derived the MSA profile for the request messages of each cluster, the next step is to extract the common features from the MSA profile into a single character sequence, which we call the \emph{request consensus prototype}, to facilitate efficient runtime comparison with an incoming request message from the system under test. 



\begin{figure*}[ht]
     \centering
     \includegraphics[width=15cm, height=3cm]{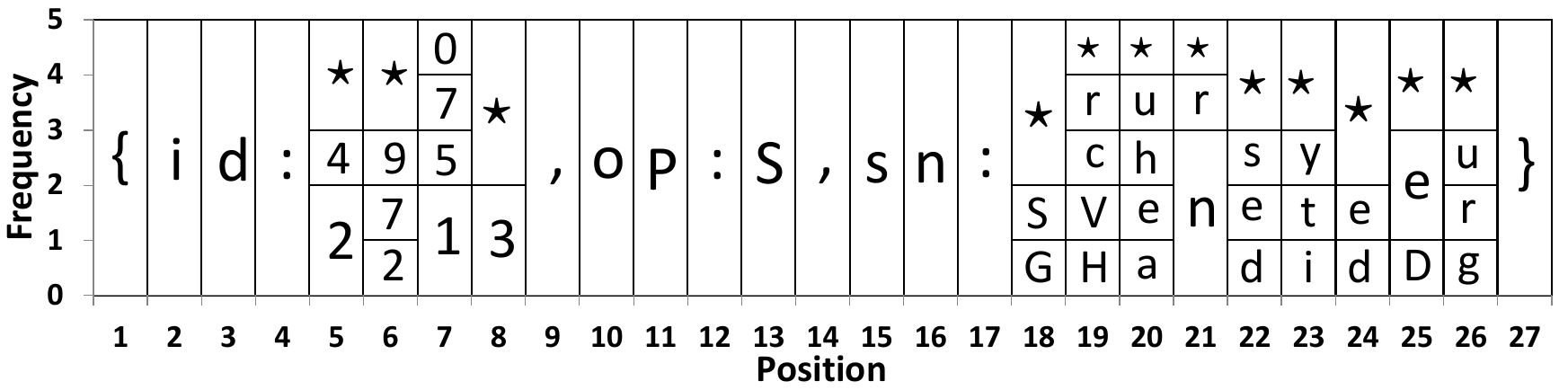}
     \caption{Character Frequencies in the Alignment Result of Search Requests.}
     \label{chap7tab:charfreq}
\end{figure*}

From all the aligned request messages in a cluster, we derive a byte (or character) occurrence count table.
Figure~\ref{chap7tab:charfreq} graphically depicts byte frequencies at each position for the example alignment in Figure~\ref{chap7fig:alignment}. Each column represents a position in the alignment result.  The frequencies of the different bytes which occur at each position are displayed as a stacked bar graph.

Based on the byte occurrence table, we formulate the \emph{request consensus prototype} by extending the concept of a \emph{consensus sequence}~\cite{ConsensusComparison} commonly used in summarising a MSA profile.  A consensus sequence can be viewed as a sequence of \emph{consensus symbols}, where the consensus symbol $c_i$ is the most commonly occurring byte at the position $i$. In our extension, a request consensus prototype, $\mathbf{p}$, is calculated by iterating each byte position of the MSA profile, to calculate a \emph{prototype symbol}, $p_i$, at each position, according to equation~\ref{eq:consensus}.

\begin{equation}
\label{eq:consensus}
    p_i = \left\{ \begin{array}{rl}
                 c_i &\mbox{ if $q(c_i) \geq f \land c_i \neq \gap$} \\
                 \perp &\mbox{ if $q(c_i) \geq \frac{1}{2} \land c_i = \gap$} \\
                 {\wildcard} &\mbox{ otherwise }
    \end{array}
        \right.
\end{equation}

Where $q(c_i)$ denotes the relative frequency at position $i$ of the consensus symbol $c_i$, $f$ is the relative frequency threshold, `$\gap$' denotes a gap, `$\wildcard$' is the `wildcard' symbol and `$\perp$' represents a truncation. After calculating the prototype symbol for each position, any truncation symbols are then deleted from the request consensus prototype.

Introducing wildcards and truncations into the message prototypes allows us to distinguish between gaps and where there is no consensus.  If the relative frequency $q(c_i)$ is at or above the threshold $f$, then we insert the consensus symbol into our prototype (unless the consensus symbol is a gap.)  If the relative frequency is below threshold, then we insert a wildcard.  If the consensus symbol is a gap and it is in the majority, then we leave that position as empty (\ie deleted).  Wildcards allow us to encode where there are high variability sections of the message.  In our experimentation, without truncations consensus sequences became artificially long as there tended to be many gaps.  By truncating the gaps, the lengths of the prototypes become similar to the typical lengths of messages in the cluster. The consensus prototype for a cluster of request messages can differentiate stable positions from variant positions. Moreover, it can identify consensus symbols that can be utilised for matching.

Applying our request consensus prototype method, using a frequency threshold $f=0.8$, to the example clusters in tables \ref{tab:searchcluster} and \ref{tab:addcluster} yields the following results:

    \hspace{-0.2cm} \textbf{Consensus prototype for the search request cluster:}

    \hspace{0.2cm}\texttt{\footnotesize{\{id:${\wildcard}{\wildcard}{\wildcard}$,op:S,sn:${\wildcard}{\wildcard}{\wildcard}{\wildcard}{\wildcard}{\wildcard}{\wildcard}$\}}}

    \hspace{-0.2cm} \textbf{Consensus prototype for the add request cluster:}

    \hspace{0.2cm}\texttt{\footnotesize{\{id:${\wildcard}{\wildcard}{\wildcard}{\wildcard}$,op:A,sn:${\wildcard}{\wildcard}{\wildcard}{\wildcard}{\wildcard}{\wildcard}{\wildcard}{\wildcard}{\wildcard}{\wildcard}{\wildcard}{\wildcard}{\wildcard}{\wildcard}$l${\wildcard}{\wildcard}{\wildcard}{\wildcard}{\wildcard}{\wildcard}{\wildcard}$\}}}

(Note that the add prototype contains an `l' from coincidentally aligning `l's from `mobile', `Will' and `postalCode'.)

\subsection{Deriving Entropy-Based Positional Weightings}
\label{ss:weightings}

The final step of the offline analysis is to calculate weightings for each byte position
in the consensus prototype to support the runtime distance matching process for request messages.
In our service emulator, generating a response message of the correct operation type is more critical
than the contents of the message payload. Thus we give a higher weighting to
the sections of the message that likely relate to the operation type.
To do this we make use of the observation that structure information (such as operation type)
is more stable than payload information.  We use entropy as a measurement of variability,
and use it as the basis to calculate a weighting for each byte position 
of the consensus prototype.

Using the MSA profile for a cluster (from section~\ref{ss:msa}), we calculate the entropy
for each column using the Shannon Index \cite{shannon:1948}, as given by equation~\ref{eq:shannon}.
\begin{equation}
\label{eq:shannon}
E_i = -\sum^{R}_{j=1} q_{ij} \log q_{ij}
\end{equation}
Where $E_i$ is the Shannon Index for the $i$th column, $q_{ij}$ is the relative frequency of the $j$th character in the character set at column $i$ and $R$ is the total number of characters in the character set.

Since we wish to give a high weighting to stable parts of the message,
for each column we invert the entropy by applying a scaling function of the form given
in equation~\ref{eq:scaling}
\begin{equation}
\label{eq:scaling}
w_i = \frac{1}{(1 + bE_i)^c}
\end{equation}
Where $w_i$ is the weighting for the $i$th column, $E_i$ is the entropy of the $i$th column, and $b$ and $c$ are positive constants.  The higher the values of $b$ and $c$, the more higher entropy columns are deweighted.  In our experiments we found the best results were obtained with $b=1$ and $c=10$.  This allows structural information to strongly dominate in the matching process.  Payload similarity is only used as a `tie-breaker'.

Columns that correspond to gaps removed from the consensus prototype are also dropped
from the weightings array.
Table~\ref{tab:weighting} gives an example weightings array for the search request consensus prototype.

\newcolumntype{C}[1]{>{\centering}m{#1}}
\newcolumntype{R}[1]{>{\raggedright}m{#1}}
\begin{table*}
\begin{tabular}{|l|c|c|c|c|C{0.4cm}|C{0.4cm}|C{0.4cm}|c|c|c|c|c|c|c|c|c|c|c|C{0.4cm}|C{0.4cm}|C{0.4cm}|C{0.4cm}|C{0.4cm}|c|}
\hline
$\mathbf{p}$ & \{ & i & d & : & ${\wildcard}$ & $\wildcard$ & $\wildcard$ & , & o & p & : & S & , & s & n & : & $\wildcard$ & $\wildcard$ & $\wildcard$ & $\wildcard$ & $\wildcard$ & $\wildcard$ & $\wildcard$ & \} \\
\hline
$\mathbf{E}$ & 0  & 0 & 0 & 0 & 1.05           & 1.33            & 1.33            & 0 & 0 & 0 & 0 & 0 & 0 & 0 & 0 & 0 & 1.61             & 1.61             & 0.95           & 1.33            & 1.33            & 1.05            & 1.33            & 0 \\ \hline
$\mathbf{w}$ & 1  & 1 & 1 & 1 &$\frac{1}{1342}$& $\frac{1}{4760}$&$\frac{1}{4760}$ & 1 & 1 & 1 & 1 & 1 & 1 & 1 & 1 & 1 &$\frac{1}{14638}$ &$\frac{1}{14638}$ &$\frac{1}{796}$ &$\frac{1}{4760}$ &$\frac{1}{4760}$ &$\frac{1}{1342}$ &$\frac{1}{4760}$ & 1 \\[0.12cm]
\hline
\end{tabular}
\caption{Example weightings for search consensus prototype calculated from the MSA profile in Figure~\ref{chap7tab:charfreq} (using equation~\ref{eq:scaling} constants $b = 1, c = 10$.)}
\label{tab:weighting}
\end{table*}

\subsection{Entropy Weighted Runtime Matching for Request Messages}
\label{ss:runtimematching}

At runtime, the formulated request consensus prototypes are used to match
incoming requests from the system under test.
We extend the Needleman-Wunsch algorithm to calculate the
matching distance between an incoming request and the request consensus prototype for each operation type (or cluster).
We modify the Needleman-Wunsch scoring function, $S$ (see equation~\ref{eq:wildcardnw}), by giving a special score for alignments with wildcard characters and multiplying all scores by their corresponding positional entropy weights (from equation~\ref{eq:scaling}).

\begin{equation}
\label{eq:wildcardnw}
S({p}_i,{r}_j) = \left\{ \begin{array}{rl}
		 w_i M &\mbox{ if ${p}_i = {r}_j \land {p}_i \neq \wildcard $} \\
		 w_i D &\mbox{ if ${p}_i \neq {r}_j \land {p}_i \neq \wildcard $} \\
		 w_i X &\mbox{ if ${p}_i = \wildcard $}
    \end{array}
	\right.
\end{equation}
Where ${p}_i$ is the $i$th character in the consensus prototype,
${r}_j$ is the $j$th character in the incoming request,
$w_i$ is the weighting for the $i$th column,
$M$ and $D$ are constants denoting the Needleman-Wunsch identical score and difference penalty, respectively,
and $X$ is the wildcard matching constant. In our experiments we used values $M=1$, $D=-1$ and $X=0$.

Using the modified scoring equation~\ref{eq:wildcardnw}, we apply Needleman-Wunsch to align the consensus prototype with an incoming request, calculating an absolute alignment score, $s$.

The relative distance, denoted as $d_{\mathrm{rel}}$, is calculated from the absolute alignment score to
normalise for consensus prototypes of different lengths, different entropy weights
and different numbers of wildcards.
The relative distance is in the range 0 to 1, inclusive, where 0 signifies the best
possible match with the consensus, and 1 represents the furthest possible distance.
It is calculated according to equation~\ref{eq:reldist}:

\begin{equation}
\label{eq:reldist}
d_{\mathrm{rel}}(\mathbf{p}, \mathbf{r}) =
	1 - \frac{s(\mathbf{p}, \mathbf{r}) - s_{\mathrm{min}}(\mathbf{p}) }
		     {s_{\mathrm{max}}(\mathbf{p})}
\end{equation}
Where $s_{\mathrm{max}}$ is the maximum possible alignment score for the given consensus prototype and
$s_{\mathrm{min}}$ is the minimum possible alignment score for the given consensus prototype.  These
are calculated in equations \ref{eq:maxscore} and \ref{eq:minscore}, respectively.

\begin{equation}
\label{eq:maxscore}
{s_{\mathrm{max}}(\mathbf{p})} = \sum_{i=1}^{|\mathbf{p}|} S(\mathbf{p}_i,\mathbf{p}_i)
\end{equation}

\begin{equation}
\label{eq:minscore}
{s_{\mathrm{min}}(\mathbf{p})} = \sum_{i=1}^{|\mathbf{p}|} S(\mathbf{p}_i,\varnothing)
\end{equation}
Where $\varnothing$ is a special symbol, different to all of the characters in
the consensus prototype.

The consensus prototype that gives the least distance to the incoming message is selected as the matching prototype, therefore identifying the matching transaction cluster.

As an example, suppose we receive an incoming add request with the byte sequence
\texttt{\footnotesize{
\{id:37,op:A,sn:Durand\}}}.
Aligning the request against the search consensus prototype yields:

\begin{tabular}{ll}
\emph{request:} & \texttt{\footnotesize{\{id:3*7,op:A,sn:*Durand\}}} \\
\emph{prototype:}        & \texttt{\footnotesize{\{id:???,op:S,sn:???????\}}} \\
\end{tabular}

\smallskip \noindent
Using equation~\ref{eq:reldist}, the weighted relative distance is calculated to be 0.0715.
In comparison, the add consensus prototype produces the alignment below with a relative distance of 0.068.

\begin{tabular}{ll}
\emph{request:} & \texttt{\footnotesize{\{id:37**,op:A,sn:********D*u*r*an*d***\}}} \\
\emph{prototype:}        & \texttt{\footnotesize{\{id:????,op:A,sn:?????????????l???????\}}} \\
\end{tabular}

\smallskip \noindent
Consequently, the add prototype is the closest matching, causing the add cluster to be selected.

\subsection{Response Generation through Dynamic Substitution}

The final step in our approach is to send a customised response
for the incoming request by performing some dynamic substitutions on a response message from the matched cluster.  Here we
use the symmetric field technique described in \cite{Du:2013}, where
character sequences which occur in both the request and response
messages of the chosen transaction are substituted with the corresponding characters from the live request in the generated response. We take the response from the centroid transaction \cite{Du:2013SoftMine} of the selected cluster and apply the symmetric field substitution.

In our example, the centroid transaction from the add cluster is given below. There is one symmetric field (boxed).

\begin{tabular}{ll}
\textit{request:} & \texttt{\footnotesize{\boxed{\texttt{\{id:24,op:A}},sn:Schneider,mobile:123456\}}} \\
\textit{response:} & \texttt{\footnotesize{\boxed{\texttt{\{id:24,op:A}}ddRsp,result:Ok\}}} \\
\end{tabular}

\smallskip \noindent
After performing the symmetric field substitution the final generated response
is:
\texttt{\footnotesize{\{id:37,op:AddRsp,result:Ok\}}}


\subsection{Implementation}


We implemented our proposed system using Java, following the steps in the
previous sections, using the architecture of Figure~\ref{fig:system_overview}. 
Due to commercial agreements we are unable to release the source code of our implementation.




\section{Evaluation}
\label{sec:evaluation}

We wanted to assess three key characteristics of our approach: \emph{accuracy}, \emph{efficiency} and \emph{robustness}, aiming to answer the following research questions:
\begin{enumerate}
\item \textbf{RQ1} (\textbf{Accuracy}): Having \emph{request consensus 
    prototypes} formulated at the pre-processing stage, is our
  approach able to generate accurate, protocol-conformant responses? At
  runtime, can \emph{positional weighting based matching} further
  improve the accuracy of our approach for generating such
  responses?
\item \textbf{RQ2} (\textbf{Efficiency}): Is our technique efficient enough to
  generate timely responses, even for large transaction libraries?

\item \textbf{RQ3} (\textbf{Robustness}): Given non-homogeneous clusters,
  which contain some messages of operation types different to the majority of messages,
  is our technique robust enough to generate accurate responses?

\end{enumerate}

In order to answer these questions, we applied our technique on six message
trace datasets of four case study protocols. We ran the technique on these datasets and assessed
its accuracy, efficiency and robustness for each. We then compared these
results to our previous techniques.


\subsection{Case Study Protocols and Traces}
\label{ss:casestudies}

\begin{table}[h]
\centering
\resizebox{0.49\textwidth}{!}{
\begin{tabular}{|l|c|c|c|c|}
  \hline
  Protocol & Binary/Text & Fields & \#Ops. & \#Transactions \\
  \hline \hline
  IMS & binary & fixed length & 5 & 800 \\ \hline
  LDAP & binary & length-encoded & 10 & 2177 \\ \hline
  LDAP text (1) & text & delimited & 10 & 2177\\ \hline
  LDAP text (2) & text & delimited & 6 & 1000 \\ \hline
  SOAP & text & delimited & 6 & 1000 \\ \hline
  Twitter (REST) & text & delimited & 6 & 1825 \\ \hline
\end{tabular}}
\caption{Experiment message trace datasets (available at~\cite{datasets})}
\label{tab:sampleset}
\end{table}

We applied our techniques to four real-world protocols,
\emph{IMS}~\cite{imsredbook} (a binary mainframe protocol),
\emph{LDAP}~\cite{ldap} (a binary directory service protocol),
\emph{SOAP}~\cite{soap} (a textual protocol, with an Enterprise Resource Planning (ERP) system messaging system services),
and \emph{RESTful Twitter}~\cite{twitter} (a JSON protocol for the Twitter social media service).

We have used one message trace dataset for each of these protocols.  In addition, \emph{LDAP} has two extra datasets: a dataset with textual representation converted from the binary dataset (denoted by
\emph{LDAP text (1)}), and another textual dataset that was used in
our prior work \cite{Du:2013SoftMine} (\emph{LDAP text (2)}).

We chose these protocols because: (i) they are widely used in enterprise
environments; (ii) some of them (LDAP text (2) and SOAP) were used in the
evaluation of our prior work \cite{Du:2013,Du:2013SoftMine}; (iii) they
represent a good mix of text-based protocols (SOAP and RESTful Twitter) and
binary protocols (IMS and LDAP); (iv) they use either fixed length, length encoding or
delimiters to structure protocol messages;\footnote{Given a protocol message,
  length fields or delimiters are used to convert its structure into a
  sequence of bytes that can be transmitted over the wire. Specifically, a
  length field is a number of bytes that show the length of another field,
  while a delimiter is a byte (or byte sequence) with a
  known value that indicates the end of a field.} and (v) each of them includes a diverse
number of operation types, as indicated by the \emph{Ops} column. The number
of request-response interactions for each test case is shown as column
\emph{\#Transactions} in Table~\ref{tab:sampleset}.

Our message trace datasets are available for download from our website~\cite{datasets}.


\subsection{10-fold Cross-Validation Approach and Evaluation Criteria}
\label{ss:crossvalidation}


{\em Cross-validation} \cite{devijver:1982} is a popular model validation
method for assessing how accurately a predictive model will perform in
practice. For the purpose of our evaluation, we applied the commonly used
10-fold cross-validation approach \cite{mclachlan:2004} to
all the six case study datasets.

We randomly partitioned each of the original interaction datasets into 10 groups. Of
these 10 groups, one group
is considered to be the {\em evaluation group} for testing our approach, and
the remaining 9 groups constitute the {\em training set}. This process is then
repeated 10 times (the same as the number of groups), so that each of the 10
groups will be used as the evaluation group once.

When running each experiment with each trace dataset, we applied our approach to each request message in
the {\em evaluation group}, referred to as the {\em incoming request}, to generate an
{\em emulated response}. We
recorded the time that our approach took to generate the response for each
incoming request. This was used to evaluate the runtime efficiency of our
consensus+weighting approach compared to other approaches.

Having generated a response for each incoming request, we utilised
five criteria to determine its accuracy, thereby evaluating the ability of our
approach to generate protocol-conformant responses. These five criteria are
explained as follows using an example shown in Table~\ref{tab:criteriaexample}. Consider the \emph{incoming
  request} {\texttt{\footnotesize{\{id:37,op:A,sn:Durand\}}}} with the
associated response {\texttt{\footnotesize{\{id:37,op:AddRsp,result:Ok\}}}}
in the transaction library. The emulated response is considered to be:







\begin{enumerate}
\item \textbf{identical} if its character sequence is identical to the
  recorded (or expected) response ({\cf} Example (i) in
  Table~\ref{tab:criteriaexample});

\item \textbf{consistent} if it is of the expected operation type and
  has the critical fields in the payload replicated ({\cf} Example (ii) in
  Table~\ref{tab:criteriaexample} where {\texttt{\footnotesize{{id}}}} is
  identical, but some of the other payload differs);

\item \textbf{protocol conformant} if its operation type corresponds to the
  expected response, but it differs in some payload information ({\cf} Example (iii)
  in Table~\ref{tab:criteriaexample} where both the
  {\texttt{\footnotesize{{id}}}} and {\texttt{\footnotesize{{result}}}}
  tags differ);

\item \textbf{well-formed} if it is structured correctly (that is, it
  corresponds to one of the valid response messages), but has the wrong
  operation type ({\cf} Example (iv) in Table~\ref{tab:criteriaexample} where
  the generated response is of a valid structure, but its operation type
  {\texttt{\footnotesize{{op:SearchRsp}}}} does not match the expected
  operation type {\texttt{\footnotesize{{op:AddRsp}}}}); and

  \item \textbf{malformed} if it does not meet any of the above criteria ({\cf} Example (v)
  in Table~\ref{tab:criteriaexample} where operation type
  {\texttt{\footnotesize{{op:AearchRsp}}}} in the generated response is invalid).

\end{enumerate}

We further consider a generated response to be {\em valid} if it meets one of
the first three criteria, that is, {\em identical}, {\em consistent} or {\em
  protocol conformant}. Otherwise, a generated response is considered to be
{\em invalid}.

%
%

\begin{table}[t]
\begin{center}
\begin{tabular}{|c||c|l|}
\hline
\label{identical}
\multirow{2}{*}{(i)} & Expected & \{id:37,op:AddRsp,result:Ok\} \\ \cline{2-3}
& Generated & \{id:37,op:AddRsp,result:Ok\} \\ \hline
\multirow{2}{*}{(ii)} & Expected & \{id:\textbf{\emph{37}},op:AddRsp,result:\emph{Ok}\} \\\cline{2-3}
& Generated & \{id:\textbf{\emph{37}},op:AddRsp,result:\textbf{\emph{AlreadyExists}}\} \\ \hline
\multirow{2}{*}{(iii)} & Expected & \{id:37,op:AddRsp,result:Ok\} \\\cline{2-3}
& Generated & \{id:\textbf{\emph{15}},op:AddRsp,result:\textbf{\emph{AlreadyExists}}\} \\ \hline
\multirow{2}{*}{(iv)} & Expected & \{id:37,op:\emph{AddRsp},result:Ok\} \\\cline{2-3}
& Generated & \{id:15,op:\textbf{\emph{SearchRsp}},result:Ok,gn:Miao,sn:Du\} \\ \hline
\multirow{2}{*}{(v)}& Expected & \{id:37,op:\emph{AddRsp},result:Ok\} \\\cline{2-3}
& Generated & \{id:15,op:\textbf{\emph{AearchRsp}},result:Ok,gn:Miao,sn:Du\} \\ \hline
\end{tabular}
\end{center}
\caption{Examples for accuracy criteria: (i) Identical, (ii) Consistent, (iii) Protocol conformant, (iv) Well-formed, (v) Malformed.}
\label{tab:criteriaexample}
\end{table}

\begin{table*}[t]
\centering
\begin{tabular}{|c|c|c|c|c|ccc|cc|}
  \hline
  \multirow{2}{*}{Protocol} & \multicolumn{2}{c|}{\multirow{2}{*}{Method}} & \multirow{2}{*}{Accuracy Ratio} & \multirow{2}{*}{No.} & \multicolumn{3}{c|} {Valid} & \multicolumn{2}{c|} {Invalid} \\ \cline{6-10}
&\multicolumn{2}{c|}{}& & &Identical & Consistent & Conformant & Well-formed & Malformed \\ \hline\hline

  \multirow{8}{2cm}{\centering IMS (binary) } & \multicolumn{2}{c|}{Whole Library} & 75.25\% & \multirow{8}{*}{800} &  400 & 202 & 0 & 198 & 0 \\ \cline{2-4} \cline{6-10} &\multicolumn{2}{c|}{Cluster Centroid} & 97.88\% & & 400 & 383 & 0 & 17 & 0 \\ \cline{2-4} \cline{6-10}
    & \multicolumn{1}{c|}{\multirow{3}{1.2cm}{\centering Consensus Only}} & f=0.5 & \textbf{\textit{100\%}}&& 400 & 400 & 0 & 0 & 0 \\ \cline{3-4}\cline{6-10}
  & & f=0.8 & \textbf{\textit{100\%}}&& 400 & 400 & 0 & 0 & 0 \\\cline{3-4} \cline{6-10}
  && f=1 & \textbf{\textit{ 100\%}}&& 400 & 400 & 0 & 0 & 0 \\ \cline{2-4}\cline{6-10}
  & \multicolumn{1}{c|}{\multirow{3}{1.2cm}{\centering Consensus + Weighting}} & f=0.5 & \textbf{\textit{100\%}}&& 400 & 400 & 0 & 0 & 0 \\ \cline{3-4}\cline{6-10}
  & & f=0.8 & \textbf{\textit{100\%}}&& 400 & 400 & 0 & 0 & 0 \\\cline{3-4} \cline{6-10}
  && f=1 & \textbf{\textit{100\%}}&& 400 & 400 & 0 & 0 & 0 \\ \hline

   \multirow{8}{2cm}{\centering LDAP (binary)} & \multicolumn{2}{c|}{Whole Library} & 94.12\% & \multirow{8}{*}{2177}  & 248 & 17 & 1784 & 36 & 92 \\ \cline{2-4} \cline{6-10}
   & \multicolumn{2}{c|}{Cluster Centroid} & 91.59\% && 263 & 17 & 1714 & 183 & 0 \\ \cline{2-4} \cline{6-10}
  & \multicolumn{1}{c|}{\multirow{3}{1.2cm}{\centering Consensus Only}} & f=0.5 & \textit{87.74\%} &&268 & 14 & 1628 &267 & 0 \\ \cline{3-4}\cline{6-10}
  & & f=0.8 & \textit{84.66\%}&& 264 & 14 & 1565 & 334 & 0 \\\cline{3-4} \cline{6-10}
  && f=1 & \textit{79.97\%}&& 259 & 14 & 1468 & 436 & 0 \\\cline{2-4}\cline{6-10}
  & \multicolumn{1}{c|}{\multirow{3}{1.2cm}{\centering Consensus + Weighting}} & f=0.5 & \textit{98.71\%} &&278 & 18 & 1853 &28 & 0 \\ \cline{3-4}\cline{6-10}
  & & f=0.8 & \textbf{\textit{99.95\%}}&& 278 & 18 & 1880 & 1 & 0 \\\cline{3-4} \cline{6-10}
  && f=1 & \textit{86.90\%}&& 267 & 16 & 1609 & 285 & 0 \\ \hline

   \multirow{8}{2cm}{\centering LDAP text (1) (text)} & \multicolumn{2}{c|}{Whole Library} & 100\% & \multirow{8}{*}{2177} &  1648 & 415 & 114 & 0 & 0 \\ \cline{2-4} \cline{6-10}
    & \multicolumn{2}{c|}{Cluster Centroid} & 100\% & &  811 & 1325 & 41 & 0 & 0 \\ \cline{2-4} \cline{6-10}
  & \multicolumn{1}{c|}{\multirow{3}{1.2cm}{\centering Consensus Only}} & f=0.5 & \textbf{\textit{100\%}}&& 1555 & 622 & 0 & 0 & 0 \\ \cline{3-4}\cline{6-10}
  & & f=0.8 & \textbf{\textit{100\%}}&& 1555 & 622 & 0 & 0 & 0 \\\cline{3-4} \cline{6-10}
  && f=1 & \textbf{\textit{100\%}}&& 1527 & 650 & 0 & 0 & 0 \\\cline{2-4} \cline{6-10}
  & \multicolumn{1}{c|}{\multirow{3}{1.2cm}{\centering Consensus + Weighting}} & f=0.5 & \textbf{\textit{100\%}}&& 1559 & 618 & 0 & 0 & 0 \\ \cline{3-4}\cline{6-10}
  & & f=0.8 &\textbf{\textit{100\%}}&& 1559 & 618 & 0 & 0 & 0 \\\cline{3-4} \cline{6-10}
  && f=1 & \textbf{\textit{100\%}}&& 1559 & 618 & 0 & 0 & 0 \\ \hline

   \multirow{8}{2cm}{\centering LDAP text (2) (text)} & \multicolumn{2}{c|}{Whole Library} &92.9\% & \multirow{8}{*}{1000}  &  927 & 2 & 0 & 71 & 0 \\ \cline{2-4} \cline{6-10}
   & \multicolumn{2}{c|}{Cluster Centroid} &73.4\% & &  509 & 225 & 0 & 252 & 14 \\ \cline{2-4} \cline{6-10}
  & \multicolumn{1}{c|}{\multirow{3}{1.2cm}{\centering Consensus Only}} & f=0.5 & \textbf{\textit{100\%}} && 808 & 192 & 0 & 0 & 0 \\ \cline{3-4}\cline{6-10}
  & & f=0.8 & \textbf{\textit{100\%}}&& 808 & 192 & 0 & 0 & 0 \\\cline{3-4} \cline{6-10}
  && f=1 & \textbf{\textit{100\%}}&& 808 & 192 & 0 & 0 & 0 \\ \cline{2-4} \cline{6-10}
   & \multicolumn{1}{c|}{\multirow{3}{1.2cm}{\centering Consensus + Weighting}} & f=0.5 & \textbf{\textit{100\%}} && 808 & 192 & 0 & 0 & 0 \\ \cline{3-4}\cline{6-10}
  & & f=0.8 & \textbf{\textit{100\%}}&& 808 & 192 & 0 & 0 & 0 \\\cline{3-4} \cline{6-10}
  && f=1 & \textbf{\textit{100\%}}&& 808 & 192 & 0 & 0 & 0 \\ \hline

  \multirow{8}{2cm}{\centering SOAP (text)}  & \multicolumn{2}{c|}{Whole Library} & 100\% & \multirow{8}{*}{1000} &  77 & 923 & 0 & 0 & 0 \\ \cline{2-4} \cline{6-10}
    & \multicolumn{2}{c|}{Cluster Centroid} & 100\% &&  98 & 902 & 0 & 0 & 0 \\ \cline{2-4} \cline{6-10}
  & \multicolumn{1}{c|}{\multirow{3}{1.2cm}{\centering Consensus Only}} & f=0.5 & \textbf{\textit{100\%}} && 96 & 904 & 0 & 0 & 0 \\ \cline{3-4}\cline{6-10}
  & & f=0.8 & \textbf{\textit{100\%}}&& 96 & 904 & 0 & 0 & 0 \\\cline{3-4} \cline{6-10}
  && f=1 & \textbf{\textit{100\%}}&& 96 & 904 & 0 & 0 & 0 \\ \cline{2-4} \cline{6-10}
  & \multicolumn{1}{c|}{\multirow{3}{1.2cm}{\centering Consensus + Weighting}} & f=0.5 & \textbf{\textit{100\%}} && 96 & 904 & 0 & 0 & 0 \\ \cline{3-4}\cline{6-10}
  & & f=0.8 & \textbf{\textit{100\%}}&& 96 & 904 & 0 & 0 & 0 \\\cline{3-4} \cline{6-10}
  && f=1 & \textbf{\textit{100\%}}&& 96 & 904 & 0 & 0 & 0 \\ \hline

    \multirow{8}{2cm}{\centering Twitter (REST) (text)}& \multicolumn{2}{c|}{Whole Library} & \textbf{99.56\%} & \multirow{8}{*}{1825} &  150 & 994 & 673 & 7 & 1  \\ \cline{2-4} \cline{6-10}
  & \multicolumn{2}{c|}{Cluster Centroid} &99.34\% & &  0 & 896 & 917 & 11 & 1  \\ \cline{2-4} \cline{6-10}
  & \multicolumn{1}{c|}{\multirow{3}{1.2cm}{\centering Consensus Only}} & f=0.5 & \textit{99.34\%}&& 0 & 893 & 920 & 11 & 1 \\ \cline{3-4}\cline{6-10}
  & & f=0.8 & \textit{99.34\%}&& 0 & 893 & 920 & 11 & 1 \\\cline{3-4} \cline{6-10}
  && f=1 & \textit{99.34\%}&& 0 & 893 & 920 & 11 & 1 \\ \cline{2-4} \cline{6-10}
    & \multicolumn{1}{c|}{\multirow{3}{1.2cm}{\centering Consensus+ Weighting}} & f=0.5 & \textit{99.34\%}&& 0 & 893 & 920 & 11 & 1 \\ \cline{3-4}\cline{6-10}
  & & f=0.8 & \textit{99.34\%}&& 0 & 893 & 920 & 11 & 1 \\\cline{3-4} \cline{6-10}
  && f=1 & \textit{99.34\%}&& 0 & 893 & 920 & 11 & 1 \\ \hline
\end{tabular}
\caption{Evaluation Results of Applying Consensus Sequence Prototype}
\label{tab:accuracy}
\end{table*}

\subsection{Compared Techniques}

To our knowledge there is no other work that generates application layer
responses directly from trace messages.  Our comparison is therefore with our
prior work, including the \textit{Whole Library}~\cite{Du:2013} and the
\textit{Cluster Centroid}~\cite{Du:2013SoftMine} approaches.
Given an incoming request, the \textit{Whole Library} approach searches the entire transaction library for
its closest matching request to synthesize its response(s). This approach is
effective in producing accurate responses. Experimental results revealed that
more than 90\% of generated responses were valid, as defined by the criteria in Section \ref{ss:crossvalidation}. However, it is generally too slow for real-time use.

The \textit{Cluster Centroid} approach reduces the number of searches to the number of transaction library clusters. Therefore, it can generate responses for real time use, but with less accuracy. We use these two approaches as baselines to evaluate our new technique.

\subsection{Evaluation Results}
\label{ss:results}


\subsubsection{\textbf{Accuracy (RQ1)}}
\label{ssb:effectiveness}

The accuracy evaluation is conducted to assess the capability of our approach for generating accurate responses. As discussed in Section~\ref{sec:approach}, our approach has two important features aimed at enhancing response accuracy, the \emph{request consensus prototype} combined with \emph{entropy-weighted distance calculation} at runtime. To measure the impact of these techniques we ran two separate sets of experiments, which are referred to as \emph{Consensus Only} and \emph{Consensus+Weighting}, respectively.
In addition, for both sets of experiments we tested for the best pre-defined
frequency threshold $f$, trying three different values.\footnotemark


Table~\ref{tab:accuracy} summarises the evaluation results of \emph{Consensus
  Only}, \emph{Consensus+Weighting}, and our prior work ({\ie}\emph{Whole
  Library} and \emph{Cluster Centroid}) experiments for the six test
datasets. The \textbf{Accuracy Ratio} column is calculated by dividing the
number of valid generated responses by the total number of interactions
tested. The last five columns give a more detailed breakdown of the different
categories of valid and invalid responses generated.

\footnotetext{As illustrated in Equation~\ref{eq:consensus}, a pre-defined
  frequency threshold is required to calculate the consensus sequence
  prototype. In bioinformatics, a number of investigations have been done for
  identifying the best threshold \cite{ConsensusComparison}. In our
  experiments, we selected the most popular 3 values, that is, $0.5$, $0.8$
  and $1$.}

Table~\ref{tab:accuracy} shows that the combined {\em Consensus+Weighting} approach achieves the highest accuracy overall for the datasets tested.  The combined approach achieves 100\% accuracy for four of the datasets, and 99.95\% and 99.34\% for the remaining two (LDAP binary and Twitter, respectively).  Twitter is the only case where the {\em Whole Library} approach is marginally better.

With respect to the impact for the frequency threshold $f$, the results show that allowing some tolerance (\ie $f < 1$) in the multiple sequence alignment can yield better results.  For the LDAP (binary) dataset the thresholds of $f=0.5$ and $f=0.8$ produced significantly higher accuracy than $f=1$.  For the other datasets the threshold had no impact on the accuracy.  The general conclusion appears that the results are not very sensitive to the value of the threshold for most scenarios.

An interesting result is that the {\em Consensus+Weighting} approach has a higher accuracy than the \emph{Whole Library} approach, even though the latter uses all the available data points from the trace library (for three datasets {\em Consensus+Weighting} is significantly more accurate, for two it has the same accuracy, for one it is slightly lower). The reason for the higher accuracy is that the {\em Consensus+Weighting} abstracts away the message payload information sections (using wildcards), so is less susceptible to matching a request to the wrong operation type but with the right payload information, whereas the \emph{Whole Library} approach is susceptible to this type of error (note the well-formed but invalid responses for the \emph{Whole Library} approach in Table~\ref{tab:accuracy}).

The impact of the entropy weightings can only be observed for the LDAP binary dataset.  For this test, the weightings significantly improve the accuracy results.  For the other datasets, no impact from the weightings can be observed, as the consensus sequence prototype by itself (\emph{Consensus Only}) already produces 99-100\% accuracy.

\medskip

\subsubsection{\textbf{Efficiency (RQ2)}}
\label{ssb:efficiency}

\newcommand{\resgentime}{
    \begin{tabular}{|>{\centering\arraybackslash}m{0.8cm} |c|>{\centering\arraybackslash}m{1.1cm} |>{\centering\arraybackslash}m{1.1cm} |>{\centering\arraybackslash}m{1.5cm} |>{\centering\arraybackslash}m{1cm}|}
        \hline
       &No.& Whole Library  & Cluster Centroid & Consensus +Weighting & Real System \\\hline\hline
      IMS  &800& 470.99 & 4.94 & 3.24& 518\\ \hline
      LDAP &2177& 835.91 & 2.77 & 2.88 & 28  \\ \hline
      LDAP text(1) &2177 & 1434.70 & 5.69 & 7.30 & 28 \\ \hline
      LDAP text(2) &1000& 266.30 & 2.44 & 1.63 & 28 \\ \hline
      SOAP &1000& 380.24 & 2.97 & 3.35 & 65 \\ \hline
      Twitter&1825& 464.09 & 32.86 & 36.62 & 417 \\ \hline
    \end{tabular}
}
\newcommand{\matchingtime}{
 \resizebox{0.49\textwidth}{!}{
    \begin{tabular}{|>{\centering\arraybackslash}m{0.8cm} |c|>{\centering\arraybackslash}m{0.8cm} |>{\centering\arraybackslash}m{0.6cm} |>{\centering\arraybackslash}m{0.6cm}
    |>{\centering\arraybackslash}m{0.6cm} |>{\centering\arraybackslash}m{0.6cm} |>{\centering\arraybackslash}m{0.6cm} |}

        \hline
       \multirow{2}{0.5cm}{}&\multirow{2}{0.5cm}{\centering No.}& \multicolumn{2}{c|}{\multirow{1}{*}{Whole Library}}  & \multicolumn{2}{c|}{\multirow{1}{*}{Cluster Centroid}} & \multicolumn{2}{c|}{Consensus+W} \\\cline{3-8}
       &&M&S&M&S&M&S \\\hline\hline
      \multirow{1}{0.5cm}{IMS} &800& 460.78&10.2 & 3.67 &1.27 & 2.68&0.56 \\ \hline
      LDAP &2177& 828.95 & 6.96 & 2.38 &0.39& 2.60&0.28 \\ \hline
      LDAP text(1) &2177 & 1425.23 &9.47& 4.38&1.31 & 5.67&1.63 \\ \hline
      LDAP text(2) &1000& 257.92 &8.38& 1.35 &1.09& 1.14 &0.49\\ \hline
      SOAP &1000& 372.58 &7.66& 1.92 &1.05 & 2.45 &0.9\\ \hline
      Twitter &1825& 412.98 &51.11& 1.47 &31.39& 1.67 &34.95 \\ \hline
    \end{tabular}
 }
}

\begin{table}[t]
\centering
\subfloat[Average Total Response Generation Time (ms)\label{tab:resgen}]{\resgentime}%
     \qquad \newline
\subfloat[Average Matching Time and Average Substitution Time. M represents the matching time, and S represents the substitution time. (ms)\label{tab:matching}]{\matchingtime}
    \qquad \newline
\caption{Approach Efficiency Evaluation Results}
\label{tab:efficiency}
\end{table}

Table \ref{tab:resgen} compares the average response generation time of the
{\em Consensus+Weighting} approach with the \emph{Whole Library} and
\emph{Cluster Centroid} approaches. Tests were run on an
Intel Xeon E5440 2.83GHz CPUs with 24GB of main memory available.
The times represent the average response generation time
across all of the requests in the datasets. 
Table \ref{tab:resgen} also lists the average response times of the real
services from which the original traces were recorded. In order to get a better
insight of the runtime performance of our approach, we separately measured
matching time and substitution time, results of which are presented in Table
\ref{tab:matching}.

The results show that the \emph{Consensus+Weighting} approach is very
efficient at generating responses, indeed much faster than the real services
being emulated. The response generation time is comparable to the
\emph{Cluster Centroid} approach, being faster for some datasets, slower for
others. Both of these approaches are about two orders of magnitude faster
than the \emph{Whole Library} approach. However, whereas the \emph{Cluster
  Centroid} approach trades off accuracy for speed, the
\emph{Consensus+Weighting} has both high accuracy and speed.


Comparing the matching time versus the substitution time (shown in Table~\ref{tab:matching}) we can observe that the \emph{Whole Library} approach consumes most of its time during the matching process (because a Needleman-Wunsch alignment is made with every request in the transaction library).  \emph{Consensus+Weighting} and \emph{Cluster Centroid} have greatly reduced matching times.  Twitter has unusually long substitution times, such that for the fast approaches, most time is spent performing the substitution.  This is due to the Twitter responses being very long, causing the symmetric field identification (common substring search) to become time consuming.

The \emph{Consensus+Weighting} generates responses faster than the real services being emulated.  This is crucial for supporting testing  of an enterprise system under test under realistic performance conditions (delays can be added to slow down the emulated response, but not the other way around). A major limitation of the \emph{Whole Library} approach is that it cannot generate responses in a time which matches the real services for fast services (such as LDAP).


\medskip

\subsubsection{\textbf{Robustness (RQ3)}}
\label{ssb:robustness}

\begin{table*}[htb]
\centering
\begin{tabular}{|c||c|c|c||c|c|c||c|c|c|}
  \cline{2-10}
    \multicolumn{1}{c|}{}&\multicolumn{3}{c||}{5\% Noise}& \multicolumn{3}{c||}{10\% Noise}&\multicolumn{3}{c|}{20\% Noise} \\\cline{2-10}
  \multicolumn{1}{c|}{}&\multicolumn{3}{c||}{Consensus+Weighting}&
  \multicolumn{3}{c||}{Consensus+Weighting}& \multicolumn{3}{c|}{Consensus+Weighting} \\
  \cline{2-10}
   \multicolumn{1}{c|}{}& f = 0.5 & f = 0.8 & f = 1 & f = 0.5 & f = 0.8 & f = 1 & f = 0.5 & f = 0.8 & f = 1  \\\cline{2-10}\hline
    IMS & \textit{99.38\%} & \textit{99.75\%}  & \textit{72.25\%} & \textit{95.75\%} & \textit{98.63\%}  & \textit{68.5\%} & \textit{ 99.75\%} & \textit{96.5\%} & \textit{68.13\%}  \\ \hline
    LDAP & \textit{97.01\%} & \textit{97.29\%}  & \textit{33.49\% } & \textit{87.09\%} & \textit{79.83\%} &\textit{49.93\%} & \textit{73.81\%} & \textit{63.67\%} & \textit{42.03\%} \\ \hline
   LDAP text (1)  & \textit{100\%} & \textit{100\%}& \textit{83.74\%} & \textit{100\%} & \textit{100\%} & \textit{84.20\%}  & \textit{100\%} & \textit{100\%} & \textit{86.50\%}  \\ \hline
  LDAP text (2) &\textit{100\%}&\textit{100\%}&\textit{100\%} &\textit{100\%}&\textit{100\%}&\textit{100\%}&\textit{100\%} & \textit{100\%} & \textit{100\%} \\ \hline
  SOAP &\textit{98.2\%}&\textit{98.2\%}&\textit{91.5\%}&\textit{100\%} &\textit{100\%}&\textit{98.2\%}& \textit{100\%} & \textit{100\%} & \textit{97.7\%} \\ \hline
  Twitter (REST)& \textit{99.34\%} & \textit{99.34\%}& \textit{22.41\%}& \textit{97.59\%}& \textit{95.01\%}& \textit{32.60\%} &  \textit{96.71\%}& \textit{94.13\%}& \textit{29.00\%}\\ \hline
\end{tabular}
\caption{Response Accuracy for Clusters with Noisy Data}
\label{tab:accuracynoisycluster}
\end{table*}

Our final test is to evaluate whether our \emph{Consensus+Prototype} approach
is robust in generating accurate responses when the clustering process (from Section~\ref{sec:clustering}) is
imperfect.  For this test we deliberately inject noise into our clusters, \ie
to create clusters where a fraction of the interaction messages are of
different operation types. The noise ratios tested were 5\%, 10\% and 20\%.
We repeated the experiments with different frequency thresholds ({\ie} 0.5,
0.8, and 1).

Table~\ref{tab:accuracynoisycluster} summarises the experimental results. The
results show that having a frequency threshold below 1 has a very big impact
on preserving the accuracy when the clustering is noisy.  A threshold of $f =
0.5$ gives the best accuracy.  When using this threshold, the accuracy stays
above 97\% for all datasets, when the noise ratio is 5\%.  As the noise ratio
increases to 20\%, the accuracy decreases significantly for binary LDAP, but
stays high for the other datasets.

Overall this is a good result.  For our approach to work best the clusters produced should be relatively clean, but there is tolerance for a small amount of noise.  A noise ratio of 20\% is considered very high.  Our actual clustering process produced perfect separation ({\ie} 0\% noise) of interaction messages by operation type for the six datasets tested.

\subsection{Industry Validation}

The proposed service emulation technique has been integrated into CA Technologies' commercial product: CA Service Virtualization. An earlier version of the technique~\cite{Muller_ODP:2014} was released as a new feature, named \emph{opaque data processing}, in version 8.0 of the product and has been sold to customers~\cite{theaustralian}. Opaque data processing has been used at customer sites to successfully emulate services of protocols not otherwise supported by the product. The emulated protocols included a proprietary extension of IMS and Sun's ONC/RPC protocol. The present technique is on the backlog for future versions.

\subsection{Threats to Validity}
We have identified some threats to validity which should be taken into consideration when generalising our experimental results:
\begin{itemize}
\item Our evaluation was performed on six datasets from four protocols.  Given the great diversity in message protocols, further testing should be performed on other message protocols.
\item The datasets were obtained by randomly generating client requests for services of different protocols.  Some real system interactions are likely to be more complicated than those of our datasets.  Further testing on real system interactions are warranted.
\end{itemize}





\section{Discussion and Future Work}
\label{sec:discussion}


We have developed an approach for automatically generating service responses from message traces which requires no prior knowledge of message structure or decoders or message schemas.
Our approach of using multiple sequence alignment to automatically generate consensus prototypes for the purpose of matching request messages is shown to be accurate, efficient and robust.  Wildcards in the prototypes allow the stable and unstable parts of the request messages for the various operation types to be separated.  Rather than using the prototypes directly for strict matching (such as using it as a regular expression) we instead calculate matching distance through a modified Needleman-Wunsch alignment algorithm.  Since we look for the closest matching prototype, the method is robust even if the prototypes are imperfect.  Moreover, this process can match requests which are slightly different to the prototypes, or are of different length to the prototypes.  This allows the system to handle requests which are outside of the cases directly observed in the trace recordings.  Weighting sections of the prototype with different importance based on the entropy, further improves the matching accuracy.

Our experimental results using the 6 message trace datasets demonstrate that our approach is able to automatically generate accurate responses in real time for most cases. Moreover, we can also see that our approach can also generate accurate responses from imperfect message clusters that contain a small number of messages of different operation types.

One limitation in our current approach is the lack of diversity in the responses generated.  We are working on an approach to identify common patterns of all responses in a cluster. A possible solution is to apply multiple sequence alignment to response messages, to distinguish stable positions from variable positions. The variable parts of responses could then be stochastically generated. 
To further improve the robustness of our approach to noisy clustering, we will utilise outlier detection techniques \cite{outlierdetection} to remove outliers of clusters before applying the alignment method.

Our approach does not consider the service state history in formulating responses. In practice a stateless model is sufficient in many cases. For example: (i) when the emulation target service is stateless, or (ii) when the testing scenario does not contain two equivalent requests, requiring different state affected responses, or (iii) where the testing scenario does not require highly accurate responses (\eg performance testing.) To address this limitation, an avenue of future exploration is to process mine the operation sequences to discover stateful models.

%
%

\section{Conclusion}
\label{sec:conclusion}

We have developed a new technique for automatically generating realistic response messages from network traces for enterprise system emulation environments that outperforms current approaches. We use the bioinformatics-inspired multiple sequence alignment algorithm to derive message prototypes, adding wildcards for high variability sections of messages. A modified Needleman-Wunsch algorithm is used to calculate message distance and entropy weightings are used in distance calculations for increased accuracy. Our technique is able to automatically separate the payload and structural information in complex enterprise system messages, making it highly robust. We have shown in a set of experiments with four enterprise system messaging protocols a greater than $99\%$ accuracy for the four protocols tested. Additionally they show efficient emulated service response performance enabling scaling within an emulated deployment environment.






\bibliographystyle{IEEEtran}
\bibliography{Du_ase_2015}

\end{document}